# The Design and Implementation of a National AI Platform for Public Healthcare in Italy: Implications for Semantics and Interoperability

**Roberto Reale**
Agenzia per l'Italia Digitale, Italy
`roberto@reale.info`

**Elisabetta Biasin**
KU Leuven CiTiP, Belgium
`elisabetta.biasin@kuleuven.be`

**Alessandro Scardovi**
Poste Italiane, Italy
`scardov3@posteitaliane.it`

**Stefano Toro**
Prolaw, Italy
`s.toro@prolaw.it`

**Abstract**

The Italian National Health Service is adopting Artificial Intelligence through its technical agencies, with the twofold objective of supporting and facilitating the diagnosis and treatment. Such a vast programme requires special care in formalising the knowledge domain, leveraging domain-specific data spaces and addressing data governance issues from an interoperability perspective. The healthcare data governance and interoperability legal framework is characterised by the interplay of different pieces of legislation. Data law is the first to be taken into proper account. It primarily includes the GDPR, the Data Governance Act, and the Open Data Directive. Also, the Data Act and the European Health Data Space proposals will have an impact on health data sharing and therefore must be considered as well. The platform developed by the Italian NHL will have to be integrated in a harmonised manner with the systems already used in the healthcare system and with the digital assets (data and software) used by healthcare professionals. Questions have been raised about the impact that AI could have on patients, practitioners, and health systems, as well as about its potential risks; therefore, all the parties involved are called to agree upon to express a common view based on the dual purpose of improving people's quality of life and keeping the whole healthcare system sustainable for society as a whole.

## 1. Introduction

Through its technical agency AGENAS (2022), the Italian National Health Service has published a notice for a competitive dialogue tender procedure relevant for the awarding of a contract for the detailed design, implementation, commissioning, and management of an Artificial Intelligence Platform to support primary healthcare assistance. The initiative falls under those planned in Mission 6, Component 1, of the Italian Recovery and Resilience Plan (PNRR) and aims to improve territorial healthcare assistance, optimising and integrating patient care processes throughout the national territory.

Artificial Intelligence adoption by the Italian NHS has the twofold objective of supporting and facilitating the diagnosis and treatment through:

- Services to support doctors in taking charge and in management of primary and secondary prevention with a view to protecting and maintaining health (screening, epidemiological investigations, early diagnosis);

- Support services to doctors for the management of chronic conditions (chronic diseases, functional deficits and disabilities resulting from a pathological state);

- Support services for doctors to analyse the medical history of patients and complete the diagnosis with non-binding suggestions from the AI systems;

- Support services for doctors to remotely conduct visits, integrated into the national telemedicine platform and enriched by AI-based analysis systems;

- In territorial care processes, AI can act as an enabling factor to improve continuity, access and personalization of care, ensuring greater effectiveness and efficiency of the health system. AI, with the ability to learn during normal operation and overcoming the codified rules typical of expert systems, can support the doctor by optimising and enhancing the anamnestic, diagnostic and monitoring processes, developing and systematising knowledge on the basis given by case studies.

Such a vast programme requires special care in formalising the knowledge domain, leveraging domain-specific data spaces and addressing data governance issues from an interoperability perspective.

## 2. Domain and framework

From the legal perspective, different pieces of legislation come into play when it comes to AI and data sharing within the healthcare domain ([1]). First and foremost: privacy and data protection. Privacy and data protection are two distinct fundamental rights that protect individuals' autonomy and dignity and strive to preserve the individual's personal sphere. They are enshrined in the EU's primary and secondary legislation, from the Treaty of Functioning of the European Union (TFEU), the Charter of Fundamental Rights of the EU, and the General Data Protection Regulation (GDPR). A tool entailing healthcare data processing will have to carefully regard the existing requirements foreseen by data protection legislation and the available guidance issued at the EU and national level. In the process of designing and developing a digital platform, the GDPR should be taken into proper account as it enshrines the principles of data processing (lawfulness, fairness and transparency, purpose limitation, data minimization, accuracy, storage limitation, integrity, confidentiality) as well as data protection by design ([2]).

Data protection by design requires that the controller shall, both at the time of the determination of the means for processing and at the time of the processing itself, implement appropriate technical and organizational measures designed to implement data-protection principles in an effective manner. Evaluations may also require to what extent it is appropriate to process personal data and how to ensure data minimisation while facilitating accuracy of data-driven decision making. Data protection by default means that the controller shall implement appropriate technical and organizational measures to ensure that, by default, only data strictly necessary for each specific purpose of the processing are used.

The GDPR itself will also require the consideration of fundamental security measures, such as the execution of a Data Protection Impact Assessment (DPIA) ([3]) at the very early stages of the project. Security measures could also concern data localisation and its safe storage within territories deemed to adequately protect the fundamental rights of natural persons.

Data protection requirements may be followed by further data-specific legislation, especially in the public environment. More specifically, the Open Data Directive entered into force in 2019 and established a minimum set of rules governing the re-use and the practical arrangements for facilitating the re-use of data (Biasin, 2022). The core aspects set by this Directive concern the FAIR (Findability, Accessibility, Interoperability, and Reuse) principles and transparency requirements. Member States are called upon to support the availability of research data, also through national policies and open access policies.

---

[1] This section offers a non-exhaustive list of relevant laws. Many others may include, for instance, intellectual property and liability laws.
[2] Articles 5 and 25 GDPR.
[3] Article 35 GDPR.

The Data Governance Act (DGA) complements the Open Data Directive and introduces a new set of actors that will need to be considered within the context of health data sharing (such as data intermediaries, data holders, and others). As a major challenge, the data altruism mechanisms will have to strictly comply with the lawfulness mechanisms foreseen by data protection legislation when personal data are processed, and this aspect may concern the design of the platform too. The implementation, in general, will have to consider the patients and all individuals at the very centre of protection in a respectful and lawful manner.

In addition to the existing legislation, two further data regulation proposals are expected to dramatically change the environment concerning data sharing in healthcare: the Data Act and the European Health Data Space (EHDS) proposals. Both pieces of regulation include interoperability requirements. The Data Act proposal offers essential requirements to facilitate the interoperability of data to be re-used between sectors. The EHDS proposal includes provisions related to the interoperability of certain health-related data sets. Member State will have to designate national contact points tasked with enforcing such obligations. Furthermore, Chapter III of the proposal would introduce interoperability aspects for Electronic Health Records (EHR) systems for their mandatory self-certification schemes.

Given the prominent role of AI in healthcare IT platforms, the AI Act proposal becomes relevant. The AI Act proposal is currently discussed in the final stages of the EU legislative process; therefore, many provisions are likely to change from the initial text tabled by the European Commission. The core element to be analysed is that the use of AI systems, alone or in combination with other technologies, shall consider fundamental rights and ethical standards from the early stage of the design process ([4]). Concerning interoperability, it is important to note that the text promotes the possibility that the Commission may develop initiatives, including of a sectorial nature, to facilitate the lowering of technical barriers hindering cross-border exchange of data for AI development, including on data access infrastructure, semantic and technical interoperability of different types of data (Rec. 81 AI Act proposal).

An AI-based platform or AI-based technology not considering cybersecurity requirements, especially on a large scale, would introduce a severe threat to the rights and freedoms of individuals and critical infrastructures in the national territories. Therefore, cybersecurity laws such as the NIS (now NIS2) Directive and the Cybersecurity Act shall be duly considered in the design and implementation of those platforms (Biasin and Kamenjašević, 2022). Furthermore, security and interoperability requirements are part of sector-specific legislation. AI-based technologies such as certain medical devices will have to thoroughly follow medical devices law's provisions on that – and platform building on these devices will need to integrate those in their design.

Finally, the healthcare environment is a special field in which ethics plays a key role, as healthcare providers, healthcare professionals, and other persons in the decision-making process may directly and strongly impact patients' lives and their close ones. Therefore, legal considerations will need to go hand in hand with existing ethics principles – such as those offered by the biomedical ethics principles of autonomy, justice, beneficence, and non-maleficence (Beauchamp and Childress, 2001). Ultimately, the main goal remains that of not harming the patient and ensuring better conditions for their lives and society as a whole.

---

[4] Many are the aspects that should be enlisted in this regard, such as issues of transparency, data quality, avoiding bias and discrimination – which for reasons of space were left out of scope.

## 3. Methods

The platform developed by the Italian NHS will have to be integrated in a harmonised manner with the systems already used in the healthcare system and with the software used by healthcare professionals. In particular, the platform will have to provide interfaces with the main and most widespread tools used by local health professionals, in order to acquire, without additional costs for health professionals, the reports of visits, examinations of imaging, laboratory dasta and prescriptions. In this context, the platform may include the use of techniques for natural language processing and for the consolidation and semantic labelling of textual documents.

With regard to training of the AI system, it will be necessary to acquire data from different specialised datasets. To this end, the use of data from literature sources available on international databases and of synthetic data will help avoiding data protection and ethical issues that arise when ingesting data directly from the Electronic Health Record (FSE). Following the initial training, the data flow that the platform will continue to receive (both through direct input by the medical users, and thanks to the integration of further national, regional and local flows) will be useful for further and continuous training and system improvement.

Given the complexity of the ecosystem and the heterogeneity of the data sources, data governance and master data management by design approaches must be envisaged, to ensure uniformity, accuracy, proper management, semantic consistency, and accountability of the processing, also from a FAIR perspective (Findability, Accessibility, Interoperability, and Reuse).

## 4. Implications

Deploying AI-driven approaches in healthcare presents numerous significant consequences for both citizens and governments, including public administrations and organisations operating in the healthcare sector. Over recent years, medical advancements have accelerated, increasing life expectancy globally. However, as people live longer, healthcare systems grapple with heightened service demands, escalating costs, and a workforce struggling to meet patient needs. Factors driving this demand include an aging population, evolving patient expectations, shifting lifestyle choices, and the constant cycle of innovation.

The ageing population's impact is particularly noteworthy. By 2050, one in four individuals in Europe and North America will be over 65, increasing the number of patients with complex healthcare needs. Addressing these patients is costly and necessitates a transition from an episodic care-based approach to a more proactive, long-term care management philosophy (Spatharou, 2020).

Healthcare expenditures are not keeping pace. Without significant structural and transformative changes, healthcare systems will have difficulty maintaining sustainability. Health systems also require a larger workforce. Although the global economy could generate 40 million new health-sector jobs by 2030, the World Health Organization (2016) projects a shortfall of 9.9 million physicians, nurses, and midwives globally during the same timeframe. There is a need to not only attract, train, and retain more healthcare professionals but also to ensure they allocate their time effectively, focusing on patient care.

Leveraging automation, artificial intelligence holds the potential to revolutionize healthcare and address some of these challenges. The European Parliament's (2016) definition of AI is worth recalling: "AI is the ability of a computer program to perform tasks or reasoning processes typically associated with human intelligence". In other words, AI can enhance care outcomes and boost care delivery's productivity and efficiency. It can also improve healthcare practitioners' daily lives, allowing them to dedicate more time to patient care, thus increasing staff morale and retention. AI can even expedite the introduction of life-saving treatments to the market.